\documentclass[twocolumn,showpacs,preprintnumbers,amsmath,amssymb]{revtex4}

\usepackage{graphicx}
\usepackage{dcolumn}
\usepackage{bm}

\begin{document}

\preprint{APS/123-QED}

\title{Ice XV: a new thermodynamically stable phase of ice}

\author{Christoph G. Salzmann,$^{1,\ast}$ Paolo G. Radaelli,$^2$ Erwin Mayer,$^3$ and
John L. Finney$^{\,4}$}

\affiliation{
$^1\!\!\!$ Department of Chemistry, University of Oxford, South Parks Road, Oxford OX1 3QR, UK\\
$^2\!\!\!$ Department of Physics, University of Oxford, Parks Road, Oxford OX1 3PU, UK\\
$^3\!\!\!$ Institute of General, Inorganic and Theoretical Chemistry, University of Innsbruck, Innrain 52a, 6020 Innsbruck, Austria\\
$^4\!\!\!$ Department of Physics and Astronomy, University College London, Gower Street, London WC1E 6BT, UK}

\date{\today}

\begin{abstract}
A new phase of ice, named ice XV, has been identified and its structure determined by neutron diffraction. Ice XV is the hydrogen-ordered counterpart of ice VI and is thermodynamically stable at temperatures below $\sim$130 K in the 0.8 to 1.5 GPa pressure range. The regions of stability in the medium pressure range of the phase diagram have thus been finally mapped, with only hydrogen-ordered phases stable at 0 K. The ordered ice XV structure is antiferroelectric (\textit{P}\=1), in clear disagreement with recent theoretical calculations predicting ferroelectric ordering (\textit{Cc}).
\end{abstract}

\pacs{64.60.Cn, 61.05.fm, 64.70.kt, 61.50.Ks}%

\email{christoph.salzmann@chem.ox.ac.uk}                              
                              
\maketitle

When water freezes, hydrogen-disordered phases of ice crystallize which exhibit orientational disorder of the hydrogen bonded water molecules \cite{Pauling1935,Petrenko1999}. Upon isobaric cooling, these phases are expected to undergo transitions to hydrogen-ordered phases in which the water molecules adopt the energetically most favored orientations (\textit{cf.} Fig. 1). However, because of the highly cooperative nature of molecular reorientation in ice, the sluggish kinetics of molecular reorientation at low temperatures often prevent these ordering transitions occurring, and the formation of `orientational glasses' is observed instead \cite{Suga1997,Salzmann2008}.

The region in the phase diagram at temperatures below the stability domain of ice VI is still `uncharted territory' (\textit{cf.} Fig. 1(a)); it is unknown which phase is thermodynamically stable under these conditions \cite{Petrenko1999}. Ice VI is hydrogen-disordered, and extending its region of stability down to 0 K would result in the problematic situation that a phase, with configurational entropy greater than zero, would be stable at 0 K. The transition of ice VI to a hydrogen-ordered phase upon cooling would be a way out of this dilemma. Alternatively, it could be that the ice II/VI and ice VI/VIII equilibrium lines meet above 0 K at an ice II/VI/VIII triple point below which the ordered ices II and VIII would be the stable phases \cite{Petrenko1999}. Based on unpublished neutron diffraction data, Kamb suggested the formation of a partially hydrogen-ordered ice VI phase \cite{Kamb1973}. Also, Johari and Whalley reported observing a very slow transformation upon cooling ice VI \cite{Johari1979}. However, no diffraction data were obtained in that study. Subsequent neutron diffraction measurements concluded that no significant structural changes were observed when pure ice VI was cooled to low temperatures \cite{Kuhs1984}. In agreement with this conclusion, Raman spectroscopic measurements of ice VI at low temperatures also did not reveal the features expected for hydrogen ordering \cite{Minceva-Sukarova1986}, even when the sample was doped with potassium hydroxide \cite{Minceva-Sukarova1988}, a recipe found to promote hydrogen ordering of ice I\textit{h} \cite{Tajima1982}. Recently, it was shown that doping the hydrogen-disordered ices V and XII with minute amounts of hydrochloric acid helps maintaining dynamic states to low enough temperatures so that the phase transitions to the hydrogen-ordered ices XIII and XIV could be observed \cite{Salzmann2006}.

\begin{figure}
\includegraphics[width=8.6cm]{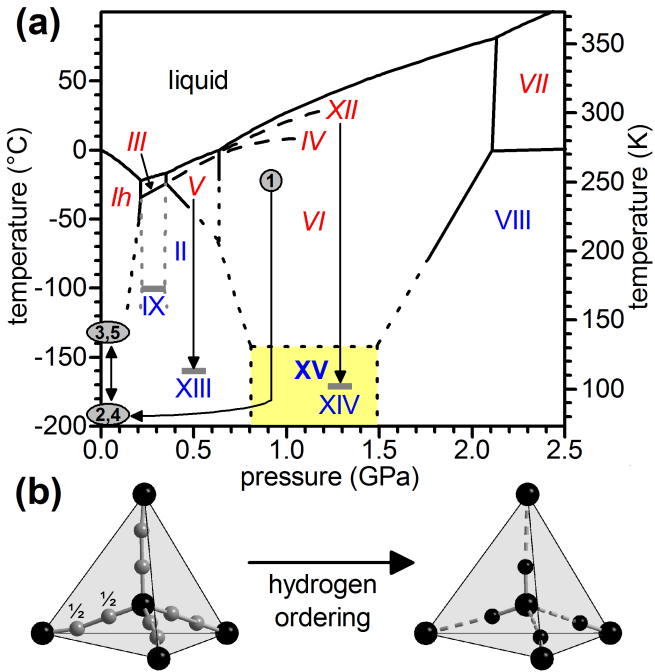}
\caption{\label{fig:fig1} (Color online) (a) The phase diagram of water and ice, including the liquidus lines of metastable ices IV and XII (long dashed lines) and extrapolated equilibria lines at low temperatures (short dashed lines). Hydrogen-disordered and ordered phases are indicated by italic and regular Roman numerals, respectively. DCl doped ice VI was produced by heating doped ice I\textit{h} to 250 K at $\sim$0.9 GPa. The consecutive \textit{p}/\textit{T} pathways are shown. (b) Schematic illustration of the hydrogen ordering process as followed by diffraction. Larger spheres indicate oxygen atoms and smaller spheres hydrogen atoms, respectively. Fully occupied atomic sites are shown in black and half occupied sites in gray.}
\end{figure}

The ordering processes in ice lead to either ferroelectric structures, with the water dipole moments adding up to yield a net moment, or antiferroelectric structures in which the water dipole moments cancel each other. The existence of ferroelectric ice would have consequences for the geochemistry and mineral physics of icy comets, moons and planets \cite{Mitra2004}. So far, one ferroelectric ice phase has been observed experimentally (ice XI) \cite{Tajima1982,Leadbetter1985}. However, computational studies have predicted that the hydrogen-ordered counterpart of ice VI should be ferroelectric as well \cite{Knight2005,Kuo2006}.

In this Letter we show that hydrochloric acid doped ice VI transforms to a new, hydrogen-ordered phase of ice, named ice XV, upon cooling. The crystal structure of ice XV is determined by neutron diffraction and compared to the computational predictions \cite{Knight2005,Kuo2006}. Finally, the thermodynamic stability of ice XV is analysed, and the revised phase diagram of water and ice is presented.

DCl doped D$_2$O ice VI samples were prepared by freezing a 0.01 mol L$^{-1}$ solution and transferring the finely ground sample into a Paris-Edinburgh pressure cell \cite{Besson1992} which was precooled to liquid nitrogen temperature, then pressurized and heated to 250 K at about 0.9 GPa. The crystallization of ice VI was followed \textit{in-situ} by recording powder neutron diffraction patterns on the PEARL beamline at ISIS. The sample was then slow-cooled to 80 K at 0.2 K min$^{-1}$ under pressure and the pressure subsequently released (\textit{cf.} Fig. 1(a)). After this, the sample was recovered from the pressure cell for further analysis on the GEM diffractometer.

Figure 2(a) shows the changes of the lattice constant \textit{c} upon heating the sample at ambient pressure from 80 K to 138 K. The lattice constant starts to contract at about 124 K upon first heating. This change was found to be reversible upon slow-cooling to 80 K at 0.1 K min$^{-1}$, with \textit{c} expanding discontinuously starting at about 130 K. Subsequent heating showed again reversible behavior, and \textit{c} reached the value obtained after the first heating. The stepwise change upon second heating was almost three times larger than observed upon first heating. It was recently shown that hydrogen order / disorder phase transitions in ices may be accompanied by such discontinuous changes in the lattice constants and that the magnitude of a step correlates with the degree of order in the hydrogen-ordered phase \cite{Salzmann2006,Salzmann2007}. The behaviour shown in Fig. 2(a) therefore suggests that a hydrogen order / disorder phase transition may be occurring.

\begin{figure}
\includegraphics[width=8.6cm]{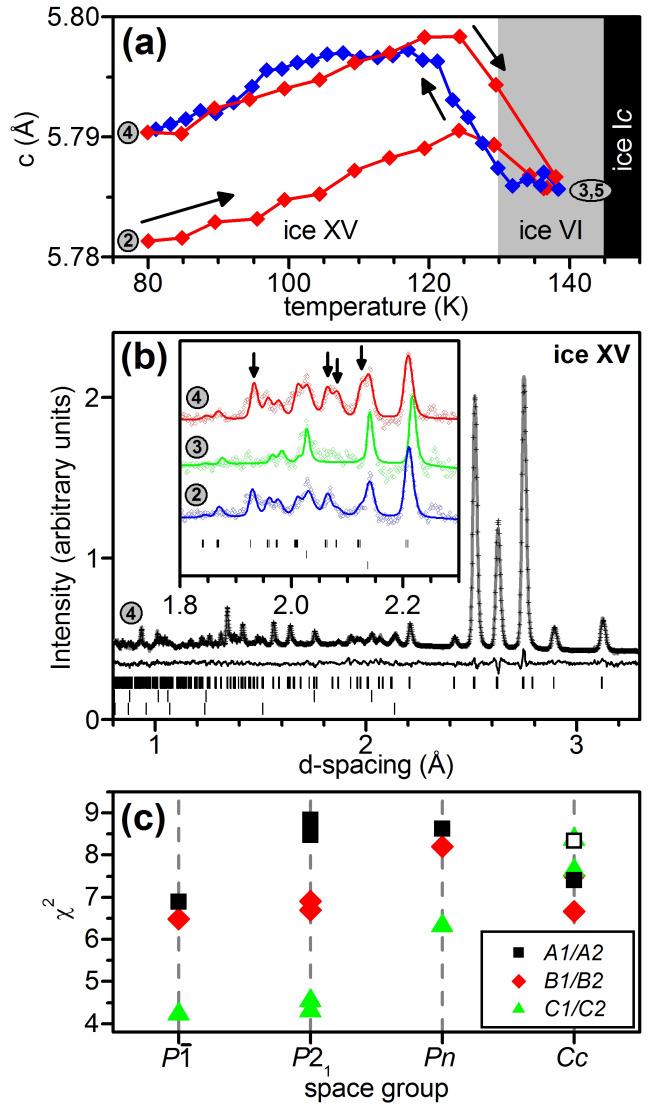}
\caption{\label{fig:fig2} (Color online) Analysis of the powder neutron diffraction data (GEM instrument). (a) Changes of lattice constant \textit{c} upon heating and cooling at ambient pressure. The numbers in (a) and (b) correspond to the \textit{p}/\textit{T} waypoints in Fig. 1(a). (b) Observed, calculated and difference neutron powder diffraction profiles for ice XV after slow-cooling to 80 K at ambient pressure (R$_{wp}$ = 0.0155, R$_p$ = 0.0186). The crystal structure was refined using the GSAS program \cite{Larsen1985} and space group \textit{P}\={1}. Linear constraints and chemical composition restraints were used to maintain the stoichiometries of the hydrogen bonds and water molecules, respectively. Tick marks indicate the positions of the Bragg peaks for ice XV, Ni and V (top to bottom). (inset) Magnification of a region where new Bragg peaks appeared. The data shown were recorded at 80 K after slow-cooling under pressure (bottom), at 138 K after heating at ambient pressure (middle) and at 80 K after slow-cooling at ambient pressure (top). Arrows indicate the positions of new Bragg peaks. (c) $\chi^2$ values obtained after converged Rietveld refinements using the fully ordered models shown in Fig. 3(b) and space group symmetries \textit{Cc}, \textit{Pn}, \textit{P}2$_1$ and \textit{P}\={1}. The $\chi^2$ value obtained for the structure predicted in refs \cite{Knight2005,Kuo2006} is depicted as an open square.}
\end{figure}

The inset in Fig. 2(b) shows the powder diffraction data recorded between the various heating and cooling steps. For the sample after slow-cooling under pressure, Bragg peaks (indicated by arrows) were observed which are either not permitted by the space group symmetry of ice VI,  \textit{P}4$_2$/\textit{nmc}, or which have zero intensity in that structure. These peaks disappeared upon heating to 138 K and reappeared upon slow-cooling to 80 K. In fact, after slow-cooling at ambient pressure, the new peaks were more intense than after slow-cooling under pressure. In agreement with the changes observed of the lattice constant \textit{c} (\textit{cf.} Fig. 2(a)), this further suggests the formation of a more highly hydrogen-ordered structure after slow-cooling at ambient pressure.

The tetragonal unit cell of ice VI comprises ten water molecules belonging to two independent but interpenetrating hydrogen bonded networks (\textit{cf.} Fig. 3(a)) \cite{Petrenko1999}. Hence, the ice VI structure can be regarded as a `self-clathrate' \cite{Kamb1965,Kuhs1984}. The central structural motifs of each network are `water hexamers' which share corners along the crystallographic \textit{c} direction and are hydrogen bonded to each other along the \textit{a} and \textit{b} directions. The \textit{P}4$_2$/\textit{nmc} space group symmetry requires ice VI to be fully hydrogen disordered, which means that the fractional occupancies of the hydrogen positions are restricted to $\frac{1}{2}$ \cite{Kamb1965,Kuhs1984}. Consequently, any change in hydrogen order has to be associated with breaking of the space group symmetry \cite{Kamb1973}.

\begin{figure}
\includegraphics[width=8.6cm]{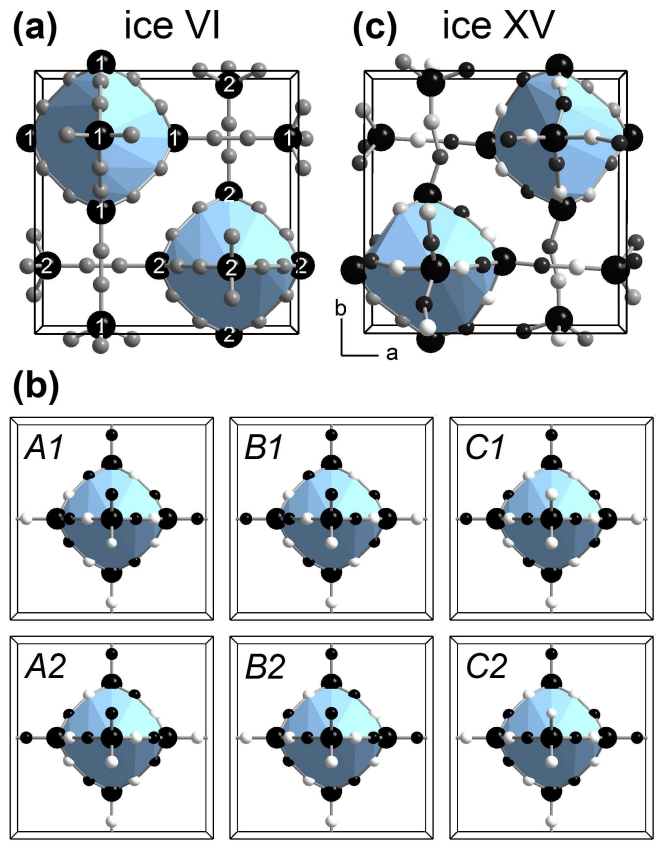}
\caption{\label{fig:fig3} (Color online) Unit cell projections of (a) hydrogen-disordered  ice VI (\textit{P}4$_2$/\textit{nmc}), (b) the six possible hydrogen-ordered structures of a single hydrogen bonded network (\textit{P}1) and (c) hydrogen-ordered ice XV (\textit{P}\={1}). The atomic site occupancies are indicated by a grayscale gradient from black (fully occupied) to white (not occupied), and the polyhedra show the front faces of the hexameric units. The oxygen atoms belonging to different networks are marked with `1' and `2' in (a).}
\end{figure}

For a single hydrogen bonded network there are six different ways in which the structure can become hydrogen ordered as shown in Fig. 3(b). Due to the chirality of the ordered hexameric units, the structures can be grouped into pairs which are related by mirror symmetry (\textit{A1/A2}, \textit{B1/B2} and \textit{C1/C2}). The \textit{A1/A2} and \textit{B1/B2} structures are polarized in the \textit{a-b} plane only, whereas the \textit{C1/C2} structures show additional polarization in the \textit{c} direction. The space groups which can accommodate two symmetry-related, fully ordered networks are \textit{Cc}, \textit{Pn}, \textit{P}2$_1$ and \textit{P}\={1}. Complete cancelation of the polarities of the two networks, and thus antiferroelectrity, is only achieved in \textit{P}\={1}. The other space group symmetries yield ferroelectric structures. Space group \textit{Cc} can be excluded conclusively due to the presence of a new Bragg peak at 1.93 \AA\ which is forbidden in \textit{Cc} (\textit{cf.} inset in Fig. 2(b)). Similarly, the new weak peak at 2.08 \AA\ cannot be indexed in the \textit{Pn} space group.

Our powder diffraction data collected after slow-cooling at ambient pressure were initially fitted by performing Rietveld refinements using the different possible fully ordered structural models. The $\chi^2$ values obtained after convergence of the refinements indicate the quality of the fits and are shown in Fig. 2(c). Consistent with the absence of the new Bragg peaks, the structural models with \textit{Cc} and \textit{Pn} space group symmetries produce unsatisfactory fits. For the \textit{P}2$_1$ and \textit{P}\={1} models, a clear preference emerges for ordered structures containing the \textit{C1} or \textit{C2} enantiomers, either as a true racemate (\textit{P}\={1}) or as a racemic mixture (\textit{P}2$_1$). For the fully ordered structures, the best fits in \textit{P}2$_1$ and \textit{P}\={1} are of similar quality; however, \textit{P}\={1} yields a significantly better agreement with the data when the occupancies of the hydrogen atom sites are refined. For the \textit{P}2$_1$ structures, the occupancies of some of the hydrogen positions increased above 1 or fell below 0 during the refinement while others converged close to $\frac{1}{2}$, behavior which indicates incorrect structural models. We therefore assign \textit{P}\={1} space group symmetry to the new hydrogen-ordered phase. Unlike \textit{P}2$_1$, space group \textit{P}\={1} can be obtained by a single irreducible representation of the high-symmetry space group \textit{P}4$_2$/\textit{nmc}. Therefore, Landau theory \cite{Landau1958,Franzen1990} permits a one-step, weakly first order transition such as the one observed here, where the order parameter is a scalar field representing the departure from half occupancy of the hydrogen sites. Out of all the possible ordered structures, the experimental structure contains the most polarized networks (\textit{C1} and \textit{C2}) and its space group symmetry allows for the most effective canceling of the network polarities.

On the basis of the reduced space group symmetry, and the reversible and discontinuous change in the lattice constant, we identify this as a new phase of ice to which we assign the Roman numeral XV. Table I shows the refined atomic site occupancies, which are, consistent with the changes in the lattice constant, more ordered after cooling at ambient pressure than after cooling under pressure, and therefore closer to the thermodynamic limit of complete hydrogen order. The \textit{pseudo}-orthorhombic unit cell of ice XV with slightly distorted hexameric units is shown in Fig. 3(c).

\begin{table}
\caption{\label{tab:table2}Fractional coordinates, isotropic atomic displacement parameters and fractional occupancies for ice XV (\textit{P}\=1). Data were collected at 80 K after cooling the sample at 0.1 K min$^{-1}$ from 138 to 80 K at ambient pressure. Deuterium positions with refined fractional occupancies $<\frac{1}{2}$ are omitted from table. \textit{a} = 6.2323 \AA, \textit{b} = 6.2438 \AA, \textit{c} = 5.7903 \AA, $\alpha$ = 90.06$^{\circ}$, $\beta$ = 89.99$^{\circ}$, $\gamma$ = 89.92$^{\circ}$. The occupancy values in parentheses are for the sample after cooling at 0.2 K min$^{-1}$ at $\sim$0.9 GPa.}
\begin{ruledtabular}
\begin{tabular}{cccccccc}
 atom & $x$ & $y$ & $z$ & U$_{iso}\ast 100$ & occ. \\
\hline
O1 & 0.2750 & 0.2515 & 0.2465 & 1.48 & 1 \\
O2 & 0.2351 & 0.5201 & 0.6193 & 1.48 & 1 \\
O3 & 0.2474 & 0.9667 & 0.6206 & 1.48 & 1  \\
O4 & 0.9610 & 0.2394 & 0.8750 & 1.48 & 1  \\
O5 & 0.5254 & 0.2715 & 0.8428 & 1.48 & 1  \\
D8 & 0.4507 & 0.2611 & 0.0172 & 2.28 & 0.853 (0.833) \\
D9 & 0.0279 & 0.2755 & 0.0046 & 2.28 & 0.819 (0.585) \\
D10 & 0.2616 & 0.3493 & 0.3728 & 2.28 & 0.794 (0.666) \\
D11 & 0.2349 & 0.1184 & 0.3358 & 2.28 & 0.901 (0.773) \\
D15 & 0.2844 & 0.6761 & 0.6384 & 2.28 & 0.616 (0.500) \\
D16 & 0.6784 & 0.2568 & 0.8845 & 2.28 & 0.853 (0.733) \\
D18 & 0.3679 & 0.4559 & 0.7073 & 2.28 & 0.804 (0.908) \\
D19 & 0.3866 & 0.0134 & 0.6779 & 2.28 & 0.926 (0.674) \\
D20 & 0.1413 & 0.0444 & 0.7009 & 2.28 & 0.581 (0.594) \\
D24 & 0.0434 & 0.3439 & 0.7612 & 2.28 & 0.634 (0.758) \\
\end{tabular}
\end{ruledtabular}
\end{table}

Density functional theory approaches have successfully predicted the structure of hydrogen-ordered ice XIV \cite{Tribello2006} and have produced results consistent with the experimental ice XIII structure being the energetic ground state in the appropriate \textit{p}/\textit{T} ranges \cite{Knight2008}. To our knowledge, there are two independent computational predictions of the hydrogen-ordered structure of ice XV \cite{Knight2005,Kuo2006}. Both studies propose that the hydrogen-ordered ground state is ferroelectric with \textit{A1} / \textit{A2} hydrogen-ordered networks and \textit{Cc} space group symmetry, which is in clear disagreement with the experimental structure reported here. The reasons for this disagreement are not understood at present.

Upon isobaric cooling, the increase in free energy of a phase correlates with its entropy, \mbox{$(\frac{\partial G}{\partial T})_p = -S$}. It follows that at temperatures below the ordering temperature ($\sim$130 K from Fig. 2(a)), hydrogen-ordered ice XV will have a lower free energy than ice VI \cite{SalzmannH2Oremark}. Since the ordering transition takes place in what was previously thought to be the stability region of ice VI, the lower temperature part of the ice VI region of stability is now assigned to the stability region of ice XV. The phase boundaries of ice XV with ices II and VIII can be constructed by considering the Clapeyron equation, $(\frac{\partial T}{\partial p})_{\Delta G} = \frac{\Delta V}{\Delta S}$, which requires approximately vertical phase boundaries between two hydrogen-ordered phases of different density (II/XV and XV/VIII), and approximately horizontal phase boundaries between a hydrogen-disordered and a hydrogen-ordered phase of similar density (VI/XV). The shaded area in Fig. 1(a) indicates the region in the phase diagram in which ice XV is expected to be thermodynamically more stable than any other of the currently known phases.

In summary, we have prepared a new, hydrogen-ordered phase of ice, named ice XV, and solved its crystal structure. Ice XV is antiferroelectric, thermodynamically stable at low temperatures in the 0.8 to 1.5 GPa pressure range and forms upon slow-cooling hydrochloric acid doped ice VI. The revised phase diagram is now finally consistent with only hydrogen-ordered phases of ice being the stable phases at 0 K. The crystal structure of ice XV once again underlines the importance of experimental results to serve as benchmarks for computational models of water and to test their abilities to reproduce the lowest energy states. Such studies could lead to improved theoretical descriptions of the interactions of water with itself, and with more complex inorganic and organic molecules.

We thank the ISIS Facility for access to the PEARL and GEM instruments, M. Tucker and W. Marshall for assistance with the PEARL experiment, B. Slater and G. Tribello for discussions and the Austrian Academy of Sciences for financial support through the APART program (C.G.S.).

\end{document}